\documentclass[preprint,pre,showpacs,showkeys,preprintnumbers,amsmath,amssymb,aps]{revtex4}

\usepackage{graphicx}
\usepackage{dcolumn}
\usepackage{bm}
\usepackage{color}

\begin{document}

\baselineskip 13pt

\title{Superposition rheology of shear-banding wormlike micelles}

\author{Pierre Ballesta}
\affiliation{Centre de Recherche Paul Pascal - CNRS UPR 8631, Avenue Schweitzer, F-33600 Pessac, France}

\author{M. Paul Lettinga}
\affiliation{Forschungszentrum J\"ulich, IFF, Weiche Materie, D-52425 J\"ulich, Germany}

\author{S\'ebastien Manneville}
\email{sebastien.manneville@ens-lyon.fr}
\affiliation{Centre de Recherche Paul Pascal - CNRS UPR 8631, Avenue Schweitzer, F-33600 Pessac, France\\
Present address: Laboratoire de Physique - CNRS UMR5672,\\ENS Lyon, 46 all\'ee d'Italie, 69364 Lyon cedex 07, FRANCE}

\date{\today}

\begin{abstract}
Wormlike micelle solutions are submitted to small-amplitude oscillatory shear superimposed to steady shear in the shear banding regime. By imposing a shear oscillation, the interface between high- and low-shear regions oscillates in time. A two-fluid semi-phenomenological model is proposed for superposition rheology in the shear banding regime, which allows us to extract a characteristic velocity for the interface dynamics from experiments involving only a standard rheometer. Estimates of the stress diffusion coefficient ${\cal D}$ can also be inferred from such superposition experiments. The validity of our model is confirmed by directly recording the interface displacement using ultrasonic velocimetry.
\end{abstract}


\maketitle

\section*{Introduction}

During the last decade, wormlike micelle solutions have become a model system to study the so-called ``shear banding'' phenomenon. Depending on the concentration, most of these surfactant systems constituted of long, cylindrical, semi-flexible aggregates undergo a shear-induced transition from a state of entangled, weakly oriented micelles to a state of highly aligned micelles above some critical shear rate $\dot{\gamma}_I$. Such a transition is strongly shear-thinning since the viscosity of the aligned state can be orders of magnitude smaller than the zero-shear viscosity of the system. Under simple shear and above $\dot{\gamma}_I$, the system spatially separates into coexisting bands of high and low viscosities corresponding respectively to the entangled and aligned states. As the shear rate is increased above $\dot{\gamma}_I$, the shear-induced structure progressively expands in the sample along the velocity gradient direction until the system is fully aligned at some shear rate $\dot{\gamma}_N$ (in this work I and N respectively stand for isotropic and nematic
in reference to the isotropic-to-nematic transition although the precise structure of the shear bands is still unclear). The rheological signature of shear banding is the existence of a horizontal plateau at a constant shear stress $\sigma=\sigma_c$ in the shear stress vs shear rate constitutive curve $\sigma(\dot{\gamma})$, which extends from $\dot{\gamma}_I$ to $\dot{\gamma}_N$. The present paper is restricted to the shear banding scenario described above and referred to as ``gradient banding'' in the literature. Another situation known as ``vorticity banding'' may also occur in wormlike micelles, where the system separates into bands bearing different stresses stacked along the vorticity direction, corresponding to a vertical portion in the flow curve, {\it i.e.} to a shear-thickening transition. A recent review of the specific rheological properties of wormlike micelles is available in \cite{Berret:2005}.

The first experimental evidence for a stress plateau in nonlinear rheological measurements was provided by \cite{Rehage:1991} on the CPCl--NaSal system. Further research effort established the generality of this peculiar feature on other wormlike micelle systems \cite{Berret:1994,Berret:1997,Soltero:1999}. Theoretically, shear banding was first interpreted in the framework of nonequilibrium phase transitions in liquid crystals \cite{Cates:1989,Olmsted:1990,Olmsted:1992}. Specific features of the wormlike micelles such as polymer-like behaviour and reversible breakage were then included by \cite{Spenley:1993} in connection with the nonlinear rheology of conventional polymers \cite{Cates:1993}. These two different approaches led to a theoretical debate about non-monotonic constitutive equations and shear banding seen either as a mechanical instability or as a nonequilibrium phase transition \cite{Schmitt:1995,Spenley:1996,Olmsted:1997,Porte:1997}. Theoretical and numerical works later focused on including stress diffusion to account for a unique stress selection and for the band dynamics \cite{Dhont:1999,Yuan:1999,Olmsted:2000} and on studying the effects of flow-concentration coupling \cite{Fielding:2003a} or the possible instabilities inherent to the models \cite{Fielding:2005}.

From the experimental point of view, phase separation under shear was ascertained for the first time by flow birefringence which showed the coexistence of bands of weakly oriented and highly anisotropic material in sheared CTAB solutions close to an isotropic-to-nematic equilibrium transition \cite{Cappelaere:1995,Makhloufi:1995}. Early nuclear magnetic resonance measurements confirmed the existence of inhomogeneous flows and the presence of differently sheared regions characterized by different order parameters \cite{Mair:1996,Britton:1997,Mair:1997} but it is not until recently that the simple shear banding scenario described above received full experimental validation from light scattering and particle tracking velocimetry in the CPCl--NaSal system \cite{Salmon:2003,Mendez:2003,Hu:2005}. In particular the so-called ``lever rule'' which, in strong analogy with first-order equilibrium phase transitions, gives the proportion $\alpha$ of the aligned state as a function of the shear rate $\dot{\gamma}$ along the stress plateau:
\begin{equation}
\label{eq_lr} \dot\gamma = (1-\alpha) \dot\gamma _I +\alpha \dot\gamma _N,
\end{equation}
appears as a rather robust feature provided that steady state is reached \cite{Salmon:2003,Lerouge:2004}.

Thus, although the exact nature of shear bands is still under debate, the coexistence of differently sheared bands is now well established \cite{Lopez:2006}. Most of latest work on shear banding has concentrated on the local flow dynamics during transients \cite{Lerouge:2004,Hu:2005}, on velocity and birefringence fluctuations and departures from the steady scenario described above \cite{Holmes:2003,Becu:2004,Lopez:2004,Lee:2005,Yesilata:2006}, on interface stability \cite{Lerouge:2006}, and on modelling such spatio-temporal dynamics \cite{Radulescu:2003,Fielding:2003b,Fielding:2004,Fielding:2006}.

In this paper we propose to use the parallel superposition technique introduced by \cite{Booij:1966a} to investigate shear banding in wormlike micelles and more precisely to access the dynamics of the interface between shear bands. In our opinion, the interest of superposition rheology has been overlooked in the literature. In particular, only a very limited number of papers are devoted to superposition measurements in complex fluids that show strong flow--microstructure coupling, {\it e.g.} associative polymers \cite{Tirtaatmadja:1997} or liquid crystalline polymers \cite{Grizzuti:2003}. Our aim is to show how this technique, which is available on most rheometers, can be used to access the dynamical behaviour of shear bands, without having to rely on involved techniques as described above. We first recall the principle of superposition rheology and illustrate it in the case of wormlike micelles sheared below $\dot{\gamma}_I$, {\it i.e.} in the homogeneous, entangled state. Then a two-fluid semi-phenomenological model is described for superposition rheology in the shear banding regime in the simple case of infinite parallel plates. This model is extended to account for experimental geometries, namely cone-and-plate, Couette, and Mooney-Couette geometries. The corresponding calculations are gathered in the appendix. Finally our model is probed experimentally on the well-studied wormlike micellar system CPCl--NaSal through superposition rheology and compared to direct measurements of the interface dynamics using ultrasonic velocimetry in Couette geometry. The results are further discussed and interpreted in terms of the stress diffusion coefficient ${\cal D}$, a key parameter in recent theoretical approaches of shear banding.

\section{One-fluid superposition rheology}

Superposition rheology as first introduced by \cite{Booij:1966a} is the addition of a small-amplitude oscillatory shear to a main steady shear. The oscillatory shear can be either parallel or perpendicular to the steady shear. Superposition allows one to probe the dynamical response of a shear-driven system and to generalize the notions of viscoelastic moduli to far-from-equilibrium conditions through a perturbation analysis. The properties of the ``superposition moduli'' and the relationships between ``parallel moduli'' and ``orthogonal moduli'' were discussed by \cite{Vermant:1998} and \cite {Dhont:2001} and applied to polymer solutions and colloidal suspensions, respectively. Here we focus on parallel superposition which is now available as an option on most recent commercial rheometers.

\subsection{Notations for one-fluid superposition rheology}
\label{not}

Let us first introduce the various notations for superposition rheology. In the following we shall use complex notations and assume that the shear rate reads
\begin{equation}
\dot\gamma =\dot\gamma_1 +\dot\gamma_2 e^{i\omega t}\,.
\label{gam}
\end{equation}
Both $\dot\gamma_1$ and $\dot\gamma_2$ are taken to be real and positive. If $\dot\gamma_2$ corresponds to a perturbation to the steady shear in the linear regime, the shear stress can be written as
\begin{equation}
\sigma =\sigma_1 +\sigma_2 e^{i\omega t}\,,
\label{sig}
\end{equation}
where $\sigma_1$ is real and $\sigma_2$ is the complex amplitude of the oscillatory part of the shear stress. The issue of specifying which variable is controlled and which is measured will be addressed below and further discussed in sect.~\ref{subsec_geom}. From eqs.~(\ref{gam}) and (\ref{sig}), two apparent viscosities are defined as
\begin{eqnarray}
\eta &=& \frac{\sigma_1}{\dot\gamma_1}\,,\label{etausual}\\
\eta_{\|}^* &=& \frac{\sigma_2}{\dot\gamma_2}\,.\label{etapar}
\end{eqnarray}
Since the oscillatory part of the shear is only a linear perturbation of the steady component, $\eta$ depends only on $\dot\gamma_1$ and reduces to the standard shear viscosity found when $\dot\gamma_2=0$. On the other hand the complex viscosity $\eta_{\|}^*$ depends on both $\dot\gamma_1$ and $\omega$, and allows one to explore the dynamical behaviour of the shear-driven system.

When both $\dot\gamma_1$ and $\dot\gamma_2$ tend to zero, one should recover the usual complex viscosity $\eta^*(\omega)$ so that
\begin{equation}
\lim_{\dot\gamma_1\rightarrow 0}\,\eta_{\|}^*(\omega,\dot\gamma_1) = \eta^*(\omega)\,.
\label{etacplx}
\end{equation}
Another useful limit is found by considering vanishing frequencies for a finite $\dot\gamma_1$. In that case, $\dot\gamma_2$ and $\sigma_2$ become steady perturbations so that eq.~(\ref{etapar}) reduces to $\eta_{\|}^*=\hbox{\rm d}\sigma_1/\hbox{\rm d}\dot\gamma_1$, which leads to
\begin{equation}
\lim_{\omega\rightarrow 0}\,\eta_{\|}^*(\omega,\dot\gamma_1) = \eta(\dot\gamma_1) + \dot\gamma_1 \frac{\hbox{\rm d}\eta}{\hbox{\rm d}\dot\gamma_1}(\dot\gamma_1)\,.
\label{eta0}
\end{equation}

\subsection{Conventional rheology of wormlike micelles in the low-shear regime}

\label{convrheol}

As already reported many times in the literature, semi-dilute solutions of wormlike micelles present an almost perfect Maxwellian behaviour in the linear regime \cite{Rehage:1988}. Such a striking feature was predicted and explained in terms of a reaction-diffusion model by \cite{Cates:1987}. However, at high frequencies, significant deviations from the Maxwell model may occur due to fast relaxation modes \cite{Fischer:1997,Yesilata:2006}. Thus a more thorough description of the low-shear rheology of polymer-like micelles is provided by the Oldroyd-B model \cite{Oldroyd:1953,Oldroyd:1955} whose linear complex viscosity $\eta^*(\omega)$ and nonlinear shear viscosity $\eta(\dot\gamma)$ read
\begin{eqnarray}
\eta^*(\omega ) &=& \eta_0\,\frac{1+i\omega \tau_2}{1+i\omega \tau_1}\,,\label{oldr_etacplx}\\
\eta(\dot\gamma ) &=& \eta_0\,\frac{1+(s_2\dot\gamma)^2}{1+(s_1 \dot\gamma)^2}\label{oldr_eta0}\,,
\end{eqnarray}
where $\tau_1$, $\tau_2$, $s_1$, and $s_2$ are characteristic times.

\begin{figure}
 \includegraphics{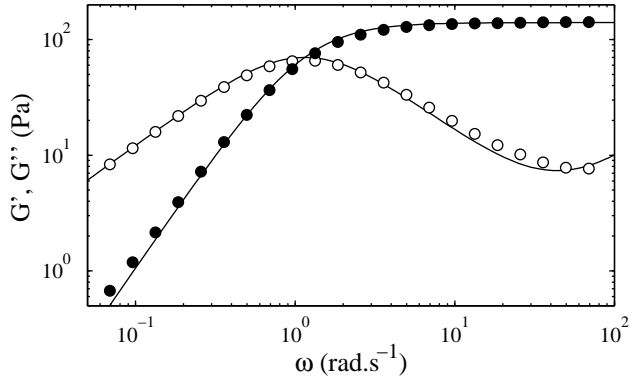}
 \caption{\label{fig1} Linear rheology of an 8$\%$~wt. CPCl--NaSal solution: storage modulus $G'$ ($\bullet$) and loss modulus $G''$ ($\circ$) versus frequency $\omega$. The solid lines correspond to an Oldroyd-B fluid (eq.~(\ref{oldr_etacplx})) with $\eta_0=122$~Pa.s, $\tau_1=0.87$~s, and $\tau_2=0.60$~ms.}
\end{figure}

In the following, we focus on a wormlike micelle solution made of cetylpyridinium chloride (CPCl, from Aldrich) and sodium salicylate (NaSal, from Acros Organics) dissolved in brine (0.5 M NaCl) with a fixed concentration
ratio [NaSal]/[CPCl]=0.5 and a total surfactant concentration of 8$\%$~wt. (unless stated differently) as described by \cite{Rehage:1988,Berret:1997}. The working temperature is $T=21^\circ$C. Figure~\ref{fig1} shows the linear viscoelastic moduli of our micellar solution measured in the Mooney-Couette geometry described below (see sect.~\ref{mooney}) with a standard stress-controlled rheometer (AR1000, TA Instruments). All the experiments in the present work were performed under controlled shear stress. Both $G'$ and $G''$ are very well described by the Oldroyd-B model (eq.~(\ref{oldr_etacplx}) with $G'+iG''=i\omega\eta^*)$ which captures the departure of $G''$ from the $\omega^{-1}$ scaling at high frequencies.

\begin{figure}
 \includegraphics{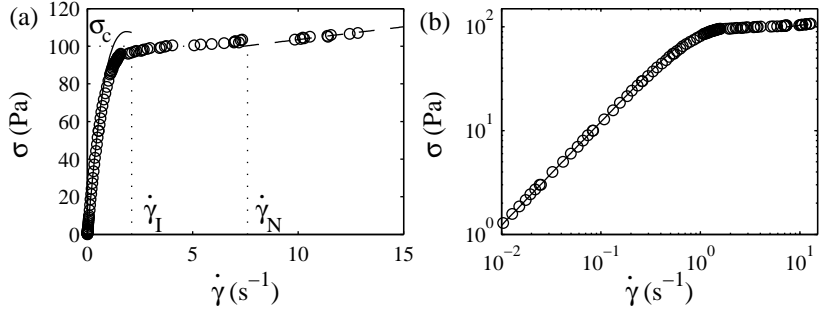}
 \caption{\label{fig2} Nonlinear rheology of an 8$\%$~wt. CPCl--NaSal solution: shear stress $\sigma$ versus shear rate $\dot\gamma$ (a) in linear scales and (b) in logarithmic scales (right). The solid line corresponds to an Oldroyd-B fluid (eq.~(\ref{oldr_eta0})) with $\eta_0=122$~Pa.s, $s_1=0.59$~s, and $s_2=0.13$~s. The dashed line is the best fit of the high-shear branch by a Bingham fluid $\sigma=\sigma_B+\eta_B\dot\gamma$ with $\sigma_B=91.7$~Pa and $\eta_B=1.13$~Pa~s. The shear banding regime extends from $\dot\gamma_I\simeq 2.2\pm 0.2$~s$^{-1}$ to $\dot\gamma_N\simeq 7.4\pm 0.4$~s$^{-1}$.}
\end{figure}
The constitutive curve $\sigma$ vs $\dot\gamma$ of the same micellar solution is shown in fig.~\ref{fig2}. As expected the fluid is weakly shear-thinning below $\dot\gamma_I\simeq2.2$~s$^{-1}$. Above $\dot\gamma_I$ very strong shear-thinning is observed and the stress saturates at a plateau value $\sigma_c\simeq 100$~Pa. This corresponds to the shear banding transition. The solid line in fig.~\ref{fig2} shows that the nonlinear rheological behaviour of our fluid in the low-shear regime is rather well captured by the Oldroyd-B model (eq.~(\ref{oldr_eta0})).

\subsection{Superposition rheology of wormlike micelles in the low-shear regime}
\label{sec_oneflu}

The superposition rheology of an Oldroyd-B fluid was computed by \cite{Booij:1966b} and leads to
\begin{equation}
\frac{\eta_{\| }^* (\omega ,\dot\gamma_1 )}{\eta_0} = \frac{1-\tau_1 \tau_2 \omega^2 (1+s_1^2 \dot\gamma_1^2 )+ (3 s_2^2-s_1^2 +s_1^2 s_2^2 \dot\gamma_1^2 )\dot\gamma_1^2+i\omega \left[\tau_1 +\tau_2 + \left(\tau_1 s_2^2+\tau_2 s_1^2\right)\dot\gamma_1^2 \right]}{\left( 1+s_1^2 \dot\gamma_1^2 \right) \left[\left(1+i \omega\tau_1 \right)^2+s_1^2\dot\gamma_1^2 \right]}\\
\,.\label{eq_annexe_oldroy_eta2}
\end{equation}
It is easily checked that eqs.~(\ref{etacplx}) and (\ref{eta0}) are recovered from eq.~(\ref{eq_annexe_oldroy_eta2}) when the limits $\dot\gamma_1\rightarrow 0$ and $\omega\rightarrow 0$ are considered.

\begin{figure}
 \includegraphics{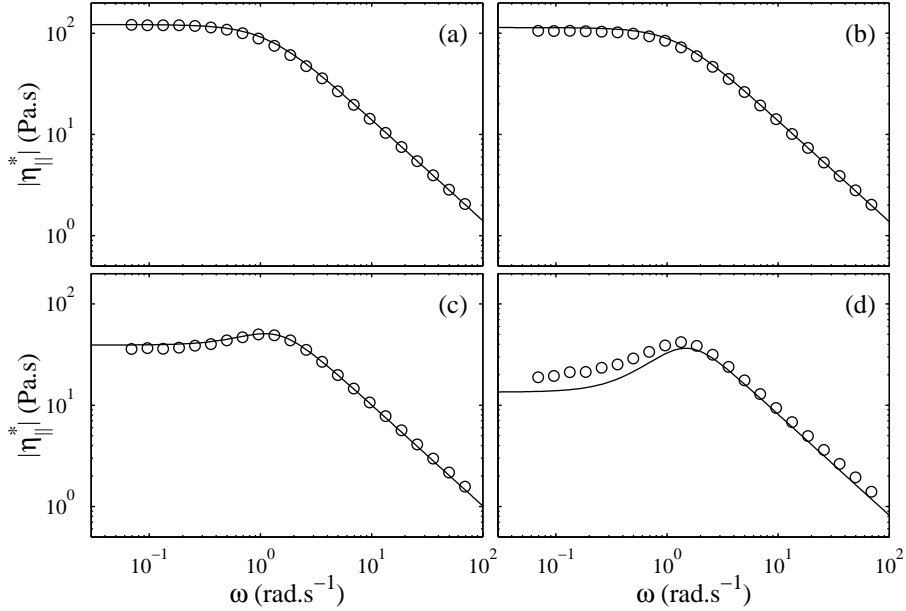}
 \caption{\label{fig3} Superposition rheology of an 8$\%$~wt. CPCl--NaSal solution in the low-shear regime: $\mid\eta_{\| }^*(\omega,\dot\gamma_1)\mid$ versus $\omega$ for (a) $\dot\gamma_1=0$, (b) 0.025, (c) 1.09, and (d) 1.49~s$^{-1}$. The solid lines correspond to an Oldroyd-B fluid (eq.~(\ref{eq_annexe_oldroy_eta2})) with $\eta_0=122$~Pa.s, $s_1=0.59$~s, $s_2=0.13$~s, $\tau_1=0.87$~s, and $\tau_2=0.60$~ms.}
\end{figure}

Figure~\ref{fig3} presents superposition measurements in the low-shear regime. The steady-state shear rates $\dot\gamma_1$ indicated in the caption of fig.~\ref{fig3} (and later figs.~\ref{fig3b}, \ref{fig3c}, and \ref{fig3d}) are the values measured by the rheometer. In all our experiments, the amplitude of the oscillatory part of the shear stress is fixed to $\sigma_2=0.5$~Pa, except for sect.~\ref{usv} where $\sigma_2=1$~Pa.
Figure~\ref{fig3} clearly shows that the four parameters inferred from figs.~\ref{fig1} and \ref{fig2} yield a good description of $\eta_{\| }^*(\omega,\dot\gamma_1)$ for all $\dot\gamma_1\lesssim 1.5$~s$^{-1}$ when used in eq.~(\ref{eq_annexe_oldroy_eta2}).
 
\section{Two-fluid superposition rheology: theoretical predictions}
\label{model}

The above results obtained in the low-shear regime prompt us to use the superposition technique in the shear banding regime. Indeed the CPCl--NaSal system is known to separate into weakly and highly sheared bands as described in the introduction \cite{Berret:1997,Porte:1997}. Previous work has shown that the shear banding phenomenon is rather simple in this particular system: the proportion of shear-induced structure is given by the lever rule (\ref{eq_lr}) and no wall slip is detected \cite{Salmon:2003,Hu:2005}. In the case of our 8$\%$~wt. CPCl--NaSal solution, the shear banding transition occurs for $\dot\gamma>\dot\gamma_I\simeq 2.2$~s$^{-1}$ and $\sigma=\sigma_c\simeq 100$~Pa. The value for the critical shear stress is in good agreement with the prediction $\sigma_c=0.67 G_0$, with $G_0=\eta_0/\tau_1$ the plateau modulus \cite{Spenley:1993}. Note however that the stress plateau is not perfectly flat at $\sigma_c$ in fig.~\ref{fig2}. This is most probably due to the curvature of the Mooney-Couette geometry which induces a significant slope of the constitutive curve in the shear banding regime \cite{Radulescu:2000,Salmon:2003}. Such a slope may also arise from flow-concentration coupling \cite{Schmitt:1995,Olmsted:1997,Fielding:2003a}. However in the absence of clear experimental evidence for such a mechanism in the literature on the system under study, we shall not refer to concentration coupling effects thereafter. In any case, the slope in the flow curve makes it hard to distinguish between the ``stress plateau'' and the homogeneous high-shear regime where the system is fully aligned (see also appendix \ref{subsec_co}).

\begin{figure}
 \includegraphics{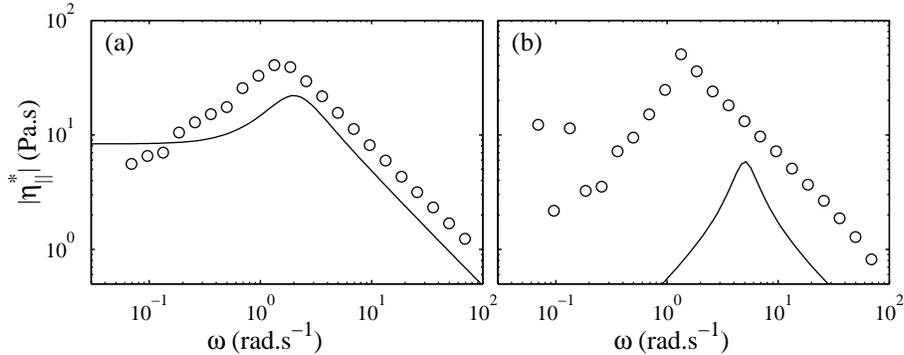}
 \caption{\label{fig3b} Superposition rheology of an 8$\%$~wt. CPCl--NaSal solution in the shear banding regime: $\mid\eta_{\| }^*(\omega,\dot\gamma_1)\mid$ versus $\omega$ for (a) $\dot\gamma_1=2.5$ and (b) 7.14~s$^{-1}$. The solid lines correspond to an Oldroyd-B fluid (eq.~(\ref{eq_annexe_oldroy_eta2})) with $\eta_0=122$~Pa, $s_1=0.59$~s, $s_2=0.13$~s, $\tau_1=0.87$~s, and $\tau_2=0.60$~ms.}
\end{figure}

Figure~\ref{fig3b} shows that the Oldroyd-B model used in the low-shear regime completely fails in describing the complex viscosity $\eta_{\| }^*(\omega,\dot\gamma_1)$ when $\dot\gamma_1>\dot\gamma_I$. This is a strong indication that the system enters the shear banding regime and that the model for superposition needs to be modified. In the following, we discuss a two-fluid model for superposition rheology in the presence of shear banding. This simple model is presented for various geometries, from the most simple geometry (infinite parallel plates) to the more complicated one actually used in our experiments (Mooney-Couette geometry). The detailed calculations for experimental geometries are presented in the appendix.

\subsection{Infinite parallel plates}
\label{ipp}

Let us first consider the case of two unbounded parallel plates in translation separated by a gap $e$. We assume that the fluid separates into bands of ``isotropic'' (I) and ``nematic'' (N) material. In a superposition experiment in the shear banding regime, the steady component of the shear stress is fixed to $\sigma_1=\sigma_c$ and the steady component of the shear rate in the isotropic (resp. nematic) material is simply $\dot\gamma_{I1}=\dot\gamma_I$ (resp. $\dot\gamma_{N1}=\dot\gamma_N$), where $\dot\gamma_I$ and $\dot\gamma_N$ are the limits of the stress plateau. This leads us to generalize the notations introduced in sect.~\ref{not} to the two-fluid case:
\begin{eqnarray}
\label{eq_sup_5} \sigma (t) &=& \sigma_c + \sigma_2 e^{i\omega t}\,, \\
\label{eq_sup_6} \dot\gamma (t) &=& \dot\gamma _1 + \dot\gamma _2 e^{i\omega t}\,, \\
\label{eq_sup_6p} \dot\gamma_I (t) &=& \dot\gamma _{I} + \dot\gamma _{I2} e^{i\omega t}\,, \\
\label{eq_sup_6pp} \dot\gamma_N (t) &=& \dot\gamma _{N} + \dot\gamma _{N2} e^{i\omega t}\,, \\
\label{eq_eta0_I}\eta_I = \frac{\sigma_c}{\dot\gamma_I}&\hbox{\rm~and~}&\eta_{\| I}^* = \frac{\sigma_2}{\dot\gamma_{I2}}\,,\\
\label{eq_eta0_N}\eta_N = \frac{\sigma_c}{\dot\gamma_N}&\hbox{\rm~and~}&\eta_{\| N}^* = \frac{\sigma_2}{\dot\gamma_{N2}}\,.
\end{eqnarray}
Since the steady shear in each phase is fixed to $\dot\gamma_I$ and $\dot\gamma_N$ respectively, $\eta_I$ and $\eta_N$ are two constants and $\eta_{\| I}^*$ and $\eta_{\| N}^*$ depend only on $\omega$. In other words $\eta_I$ and $\eta_N$ are the apparent viscosities of the isotropic and nematic materials under a steady shear stress $\sigma_1=\sigma_c$, while $\eta_{\| I}^*$ and $\eta_{\| N}^*$ correspond to the dynamical behaviours of the two phases for $\sigma_1=\sigma_c$.

In the absence of wall slip, the lever rule (\ref{eq_lr}) is a mere consequence of the continuity of the velocity at the interfaces between bands \cite{Salmon:2003} and the proportion of shear-induced structure $\alpha(t)$ obeys
\begin{eqnarray}
\label{eq_alpha_dl1}\alpha (t) &=& \alpha_1 + \alpha_2 e^{i\omega t}\,,\\
\label{eq_lr1}\dot\gamma(t) &=& (1-\alpha(t)) \dot\gamma_I(t) +\alpha(t) \dot\gamma_N(t)\,,\\
\label{eq_lr2}\dot\gamma_1 &=& (1-\alpha_1) \dot\gamma_I +\alpha_1 \dot\gamma_N\,.
\end{eqnarray}
Note that the fact that the ``instantaneous'' lever rule (\ref{eq_lr1}) applies at all times actually results from a steady-state approximation of the Navier-Stokes equations, {\it i.e.} from assuming that $\rho\partial_t v\ll\partial_z\sigma$. This assumption will be checked below {\it a posteriori}. Moreover eq.~(\ref{eq_lr2}) leads to the lever rule for the apparent viscosity
\begin{equation}
\label{eta_ipp}\frac{1}{\eta} = \frac{\dot\gamma_1}{\sigma_c} = \frac{1-\alpha_1}{\eta_{I}}+\frac{\alpha_1}{\eta_{N}}\,.
\end{equation}

Using the above notations and restricting the analysis to linear response, it is easily shown that
\begin{eqnarray}
\dot\gamma_2 = \left( 1-\alpha_1  \right) \frac{\sigma_2}{\eta^*_{\| I}} +\alpha_1 \frac{\sigma_2}{\eta^*_{\| N}} +\alpha_2 \sigma_c \left(\frac{1}{ \eta_{N}}-\frac{1}{\eta_{I}} \right) \,.\label{eq_dg1pdg2eg}
\end{eqnarray}
In order to get an expression for $\eta_{\| }^*=\sigma_2/\dot\gamma_2$ in the shear banding regime, we need to link $\alpha_2$ and $\sigma_2$. We chose to use the reaction-diffusion model proposed by \cite{Radulescu:1999} that assumes the existence of a single band and shows that the interface between the isotropic and nematic regions moves at a velocity $c$ that only depends on the difference $\sigma(r_c)-\sigma_c$, where $r_c$ is the position of the band, and vanishes for $\sigma(r_c)=\sigma_c$. More precisely, if one assumes the shear-induced structure to be located from $r=0$ to $r=r_c(t)=\alpha(t) e$, with $r$ being the coordinate across the gap, the model predicts
\begin{equation}
\frac{\hbox{\rm d}\alpha}{\hbox{\rm d}t}=
\frac{\sigma (r)-\sigma_c}{e}\,\left.\frac{\hbox{\rm d}c}{\hbox{\rm d}\sigma}\right|_{\sigma_c}
= \frac{\sigma (r)-\sigma_c}{\sigma_c}\,\frac{c_0}{e}=\frac{\sigma_2 e^{i\omega t}}{\sigma_c}\,\frac{c_0}{e}\,,\label{eq_dothn1}
\end{equation}
where we have introduced the characteristic velocity $c_0$ defined by $c_0/\sigma_c=\left.\hbox{\rm d}c/\hbox{\rm d}\sigma\right|_{\sigma_c}$. Equation~(\ref{eq_alpha_dl1}) then leads to
\begin{eqnarray}
\label{eq_alp2}
\alpha_2 = \frac{\sigma_2}{\sigma_c}\,\frac{c_0}{i\omega e}\,.
\end{eqnarray}
Experimentally the amplitude $\alpha_2 e$ of the oscillations of the interface position $r_c(t)$ should be accessible through the time-resolved velocimetry techniques mentioned in the introduction provided that the spatial resolution is fine enough. An estimate of $c_0$ from direct measurements of $\alpha_2$ using ultrasonic velocimetry will be presented in sect.~\ref{usv}. Finally inserting eq.~(\ref{eq_alp2}) into eq.~(\ref{eq_dg1pdg2eg}) and using the definition (\ref{etapar}) yields
\begin{equation}
\frac{1}{\eta_{\| }^*} =\frac{1-\alpha_1}{\eta^*_{\| I}}+\frac{\alpha_1}{\eta^*_{\| N}}+ \frac{ \dot\gamma_N-\dot\gamma_I}{\sigma_c}\,\frac{c_0}{i \omega e}\,.\label{eq_annexe_finale}
\end{equation}

Equation~(\ref{eq_annexe_finale}) shows that the complex viscosity in parallel superposition involves two terms:
\begin{eqnarray}
\frac{1}{\eta_{L}}&=&\frac{1-\alpha_1}{\eta^*_{\| I}}+\frac{\alpha_1}{\eta^*_{\| N}}\,,\label{visc_lr}\\
\frac{1}{\eta_{D_\infty}}&=&\frac{ \dot\gamma_N-\dot\gamma_I}{\sigma_c}\, \frac{c_0}{i \omega e}\,.\label{visc_dyn}
\end{eqnarray}
The first term $\eta_{L}(\dot\gamma_1,\omega)$ corresponds to the ``steady'' lever rule (\ref{eta_ipp}) applied to the complex viscosities ${\eta^*_{\| I}}$ and ${\eta^*_{\| N}}$ and depends on both $\dot\gamma_1$ (through $\alpha_1$) and $\omega$ (through $\eta^*_{\| I}$ and $\eta^*_{\| N}$). In principle ${\eta^*_{\| I}}(\omega)$ and ${\eta^*_{\| N}}(\omega)$ are accessible through superposition measurements in the homogeneous states at $\dot\gamma_1=\dot\gamma_I$ and $\dot\gamma_1=\dot\gamma_N$ respectively (or at least by extrapolation of ${\eta^*_{\|}}(\dot\gamma_1,\omega)$ when $\dot\gamma_1\rightarrow\dot\gamma_I^-$ and $\dot\gamma_1\rightarrow\dot\gamma_N^+$), so that $\eta_L(\omega,\dot\gamma_1)$ is known once $\alpha_1$ is known via eq.~(\ref{eq_lr2}).

The second term $\eta_{D_\infty}(\omega)$ accounts for the dynamics of the interface between the two bands and does not depend on $\dot\gamma_1$. Note however that this ``dynamical'' term depends on the geometry since $e$ shows up in eq.~(\ref{visc_dyn}). Since good approximations of $\dot\gamma_I$, $\dot\gamma_N$, and $\sigma_c$ are given by nonlinear rheological measurements, the only unknown in eq.~(\ref{visc_dyn}) is $c_0$. We conclude that superposition rheology in the shear banding regime should provide an experimental means of probing the dynamics of the interface between shear bands through the measurement of $c_0$.  In practice the various parameters involved in eq.~(\ref{eq_annexe_finale}) are not that easy to extract from independent measurements. As already pointed out the limits of the stress plateau are not always clear (see fig.~\ref{fig2}). But the largest difficulty probably lies in getting a good approximation for the dynamical behaviour of the shear-induced structure $\eta^*_{\| N}(\omega)$ from superposition measurements at $\dot\gamma_1\gtrsim \dot\gamma_N$. Indeed the high-shear branch of the flow curve is sometimes impossible to access due to flow instabilities that tend to expel the sample from the measuring tool at high shear rates \cite{Berret:1997,Hu:2005}. Still one may argue that $\eta^*_{\| N}(\omega)$ could also be inferred from superposition experiments in the shear banding regime by looking at the dependence of $\eta^*_{\|}(\omega,\dot\gamma_1)$ on $\alpha_1$ in eq.~(\ref{eq_annexe_finale}). We shall further discuss this point below in sect.~\ref{discuss}.

A simple way to overcome the difficulty raised by $\eta^*_{\| N}$ is to focus on the limit $\alpha_1\rightarrow 0$, {\it i.e.} just at the onset of shear  banding. In this limit
eq.~(\ref{eq_annexe_finale})  becomes
\begin{equation}
\lim_{\alpha_1\rightarrow 0}\,\frac{1}{\eta_{\| }^*} =\frac{1}{\eta^*_{\| I}}+ \frac{ \dot\gamma_N-\dot\gamma_I}{\sigma_c}\,\frac{c_0}{i \omega e}\,,\label{eq_detc0}
\end{equation}
where $c_0$ is the only unknown parameter since $\eta^*_{\| I}$ is known from measurements in the low-shear regime. Experimentally, $\alpha_1$ is varied for a given frequency $\omega$. The value at the origin of the linear regression of $1/\eta^*_{\|}$ vs $\alpha_1$ is then eq.~(\ref{eq_detc0}), from which $1/\eta_{D_\infty}(\omega)$ is determined. Finally a linear fit of $\eta_{D_\infty}(\omega)$ vs $\omega$ yields $c_0$. This fitting procedure will be tested in sect.~\ref{mooney}.

To conclude this discussion of eq.~(\ref{eq_annexe_finale}), let us check the validity of the ``instantaneous'' lever rule (\ref{eq_lr1}). In the case of oscillating velocity and stress fields, neglecting the time derivative in the Navier-Stokes equation is equivalent to setting $\rho\omega v_2\ll\sigma_2/e$ where $v_2\simeq\dot\gamma_2 e$. In the low-frequency limit, eq.~(\ref{eq_annexe_finale}) yields $\eta^*_{\|}=\sigma_2/\dot\gamma_2\simeq i\omega e\sigma_c/c_0 (\dot\gamma_N-\dot\gamma_I)$ so that the steady-state approximation holds if
\begin{equation}
\rho c_0 e\,\frac{\dot\gamma_N-\dot\gamma_I}{\sigma_c}\ll 1\,.
\label{eq_approx}
\end{equation}
With the typical values $\rho=10^3$~kg.m$^{-3}$, $c_0\lesssim 1$~mm.s$^{-1}$ (as will be checked experimentally below), $e=1$~mm, $\dot\gamma_N-\dot\gamma_I=10$~s$^{-1}$, and $\sigma_c=100$~Pa, the left-hand side of eq.~(\ref{eq_approx}) is about $10^{-4}$. Hence the approximation holds at least in the low-frequency limit of eq.~(\ref{eq_annexe_finale}). More generally, the steady-state approximation reads $\rho\omega e^2\ll\mid\eta^*_{\|}\mid$. It can be checked from figs.~\ref{fig3} and \ref{fig3b} that, for the highest frequencies achieved in our experiments ($\omega\simeq 100$~rad.s$^{-1}$), one always keeps $\mid\eta^*_{\|}\mid\gtrsim 1$~Pa.s, so that, with $\rho\omega e^2\lesssim 0.1$, the approximation remains valid.

\subsection{Experimental geometries}
\label{subsec_geom}

Standard experiments use cone-and-plate or Couette geometries (or their combination known as the Mooney-Couette geometry). The changes that the use of such geometries induces in eqs.~(\ref{eq_annexe_finale})--(\ref{visc_dyn}) are described in details in the appendix. It is shown in sect.~\ref{subsec_cp} and \ref{subsec_co} that the expressions found for $\eta^*_{\|}$ in both the cone-and-plate and the Couette geometries can be written in forms similar to eq.~(\ref{eq_annexe_finale}). 

In particular, in both cone-and-plate and infinite parallel plates, it is seen from  eq.~(\ref{eq_alp2}) that $\alpha_2$ diverges at low frequencies for fixed $\sigma_2$, {\it i.e.} under controlled stress. Such a behaviour is a direct consequence of the highly nonlinear fluid response under controlled stress in flat geometries, where jumps between the two shear branches of the flow curve are expected. Therefore, to ensure that the experiments are conducted inside the stress plateau for all frequencies, superposition rheology in the cone-and-plate geometry requires to work under controlled shear rate, so that $\alpha_2$ (and thus $\dot\gamma_2$ through eqs.~(\ref{eq_alpha_dl1})--(\ref{eq_lr2})) always remains a linear perturbation of the steady shear. Since a controlled-stress rheometer is used in the present work and since ultrasonic velocimetry is not available in the cone-and-plate geometry, we shall rather focus on the Couette geometry where the divergence of $\alpha_2$ does not occur.

More precisely, sect.~\ref{subsec_co} shows that in the ``small-gap approximation,'' {\it i.e.} when the gap $e$ is small enough compared to the radius $R_0$ of the inner cylinder, the case of a Couette geometry reduces exactly to the case of infinite parallel plates provided that $\eta_{D_\infty}$ is replaced by
\begin{equation}
\frac{1}{\eta_{D}} = \frac{ \dot\gamma_N-\dot\gamma_I}{\sigma_c}\, \frac{c_0}{i\omega e +\frac{2c_0 e}{R_0}} = \frac{1}{\eta_{D_\infty}}\,\frac{1}{1-\frac{2ic_0}{\omega R_0}}\, . \label{etaD_geom}
\end{equation}
This corresponds to the zero-order version of eq.~(\ref{eq_E_co}), {\it i.e.} it assumes that both the term of order $e/R_0$ in eq.~(\ref{etaD_co}) and the first-order corrective term $1/\eta_\partial$ given by eq.~(\ref{etad_co}) can be neglected. As seen on eq.~(\ref{eq_a2_co}) the curvature of the Couette geometry prevents $\alpha_2$ from diverging at low frequencies so that superposition measurements can be performed under controlled stress.

Finally sect.~\ref{subsec_mc} shows that the case of a Mooney-Couette geometry of height $h$ can be handled by considering the proportions $\epsilon_{co}=(1+R_0/2h)^{-1}$ and $\epsilon_{cp}=1-\epsilon_{co}$ of the surface respectively covered by the Couette ($co$) and by the cone-and-plate ($cp$) geometries relative to the total surface. In particular, $\eta_{\| }^*$ is given by the following average of the corresponding viscosities $\eta_{\| co}^*$ and $\eta_{\| cp}^*\,$:
\begin{equation}
\label{eq_e_geom} \eta_{\| }^* =\epsilon_{ co} \eta_{\| co}^*+\epsilon_{cp}\,\eta_{\| cp }^*\,.
\end{equation}
However to close the problem one has to specify the values of $\alpha_1$ in the two parts of the geometry. As shown in the appendix, this leads to serious complications and the interpretation of superposition measurements in the Mooney-Couette geometry requires in principle the full knowledge of the dynamical behaviours $\eta_{\| I}^*$ and $\eta_{\| N}^*$ of the high- and low-viscosity materials. 

To keep things analytically tractable and although this may be a crude approximation of the actual behaviour in the Mooney-Couette geometry, we shall assume that
\begin{equation}
\label{eq_e_mc_approx} \frac{1}{\eta_{\| }^*} =\frac{\epsilon_{ co}}{\eta_{\| co}^*}+ \frac{\epsilon_{cp}}{\eta_{\| cp }^*}\,,
\end{equation}
which is consistent with eq.~(\ref{eq_e_geom}) only for $e/R_0\ll 1$. Using the effective $\alpha_1$ found in sect.~\ref{subsec_mc} and given by 
\begin{equation}
\label{eq_a1_geom} \alpha_1 =\epsilon_{co} \frac{R_0}{e} \left( \sqrt{\frac{\sigma_1 -\epsilon_{cp}\,\sigma_c}{\epsilon_{co} \sigma_c}}-1 \right) +\epsilon_{cp}\, \frac{\dot\gamma_1 -\dot\gamma_I}{\dot\gamma_N -\dot\gamma_I}\,,
\end{equation}
together with eq.~(\ref{eq_E_co}) at zero order in $e/R_0$ for $\eta_{\| co}^*$ and eq.~(\ref{eq_E_cp2}) with $e=R_0\tan\beta$ for $\eta_{\| cp}^*$, one finds
\begin{equation}
\frac{1}{\eta_{\| }^*} =\frac{1-\alpha_1}{\eta^*_{\| I}}+\frac{\alpha_1}{\eta^*_{\| N}}+ \frac{ \dot\gamma_N-\dot\gamma_I}{\sigma_c}\,\frac{c_0}{i \omega e}\left(\frac{\epsilon_{co}}{1-\frac{2ic_0}{\omega R_0}}+2\epsilon_{cp}\right)\,.\label{eq_annexe_finale2}
\end{equation}
Equation~(\ref{eq_annexe_finale2}) is exactly eq.~(\ref{eq_annexe_finale}) up to a corrective frequency-dependent term on $c_0$ that accounts for the Mooney-Couette geometry.
Thus under the above assumptions we may still use the data analysis procedure described above in sect.~\ref{ipp} (see eq.~(\ref{eq_detc0})) on experimental data recorded in the Mooney-Couette geometry. 

\section{Two-fluid superposition rheology: experimental results}

\subsection{Superposition experiments in the Mooney-Couette geometry}
\label{mooney}

Superposition experiments were performed in the shear banding regime on the previous $8\%$~wt. CPCl--NaSal solution under controlled stress in a Mooney-Couette geometry with inner radius $R_0=24$~mm, outer radius $R_1=25$~mm, and height $h=30$~mm. Using the notations defined above, this corresponds to $\epsilon_{co}\simeq 0.7$ so that we can not neglect the presence of the cone. The small-gap approximation holds since $e/R_0\simeq 0.04$. A solvent trap is used to prevent evaporation and we checked that no significant change of the rheological properties of our micellar solution occurs over the $\sim 8$ hour maximal duration of our experiments.

\begin{figure}
 \includegraphics{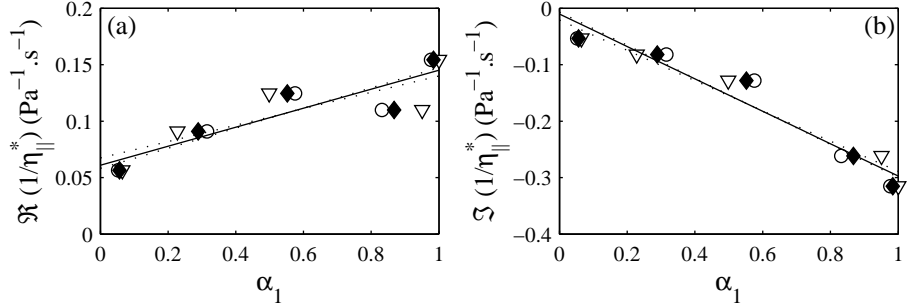}
 \caption{\label{fig4} (a) Real and (b) imaginary parts of $1/\eta_{\| }^*(\dot\gamma_1,\omega)$ versus $\alpha_1$ deduced from eq.~(\ref{eq_a1_co}) ($\circ$), from eq.~(\ref{eq_a1_cp}) ($\triangledown$), and from eq.~(\ref{eq_a1_mc}) ($\blacklozenge$). The solid lines are the best linear fits of the $\blacklozenge$ data while the dotted lines show the linear fits obtained using the $\circ$ and $\triangledown$ data. The frequency is $\omega=0.26$~rad.s$^{-1}$. The fluid under study is an 8$\%$~wt. CPCl--NaSal solution. }
\end{figure}

\begin{figure}
 \includegraphics{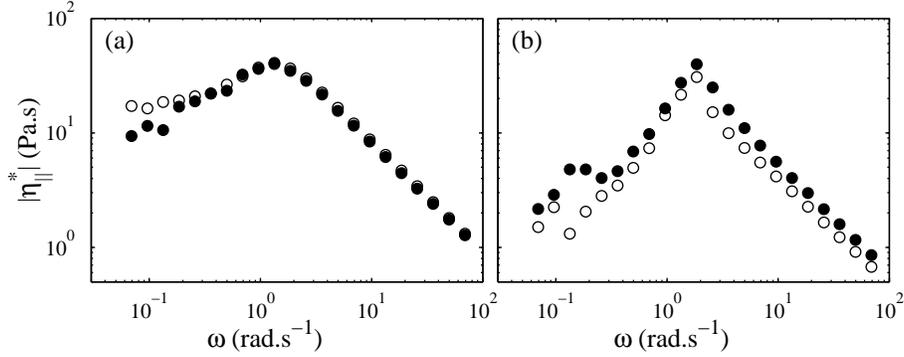}
 \caption{\label{fig3c} (a) $\mid\eta_{\| I}^*(\omega)\mid$ measured closest to the onset of shear banding for $\dot\gamma_1=1.7$~s$^{-1}\lesssim\dot\gamma_I$ ($\circ$) and inferred from the fitting procedure based on eq.~(\ref{eq_annexe_finale2}) ($\bullet$). (b) $\mid\eta_{\| N}^*(\omega)\mid$ measured in the high-shear state for $\dot\gamma_1=11.1$~s$^{-1}$ ($\circ$) and deduced from eq.~(\ref{eq_annexe_finale2}) ($\bullet$). The fluid under study is an 8$\%$~wt. CPCl--NaSal solution.}
\end{figure}

Figure~\ref{fig4} shows the experimental $1/\eta_{\| }^*(\alpha_1)$ data obtained when varying the imposed steady shear stress $\sigma_1$ ({\it i.e.} the average proportion $\alpha_1$ of oriented phase) for a given frequency $\omega$. $1/\eta_{\| }^*(\alpha_1)$ is inferred from the raw data $\eta_{\| }^*(\dot\gamma_1,\omega)$ at fixed $\omega$ (see figs.~\ref{fig3b} and \ref{fig3c} for examples of such raw data).
To test the robustness of the linear behaviour of $1/\eta_{\| }^*$ vs $\alpha_1$ expected from eq.~(\ref{eq_annexe_finale2}), the data were plotted against $\alpha_1$ computed from eq.~(\ref{eq_a1_co}) alone ({\it i.e.} taking $\epsilon_{co}=1$ and neglecting the cone-and-plate part of the geometry, see $\circ$ symbols), from eq.~(\ref{eq_a1_cp}) alone ({\it i.e.} taking $\epsilon_{cp}=1$ and neglecting the Couette part of the geometry, see $\triangledown$ symbols), and from the full eq.~(\ref{eq_a1_mc}) with $\epsilon_{co}=0.7$ and $\epsilon_{cp}=0.3$ (see $\blacklozenge$ symbols). The quality of the three linear fits are similar and the values of the slopes as well as the intercepts at $\alpha_1=0$ are all very close. We conclude that the linear behaviour predicted by eq.~(\ref{eq_annexe_finale2}) is indeed observed and that the way $\alpha_1$ is computed is not critical.

\begin{figure}
 \includegraphics{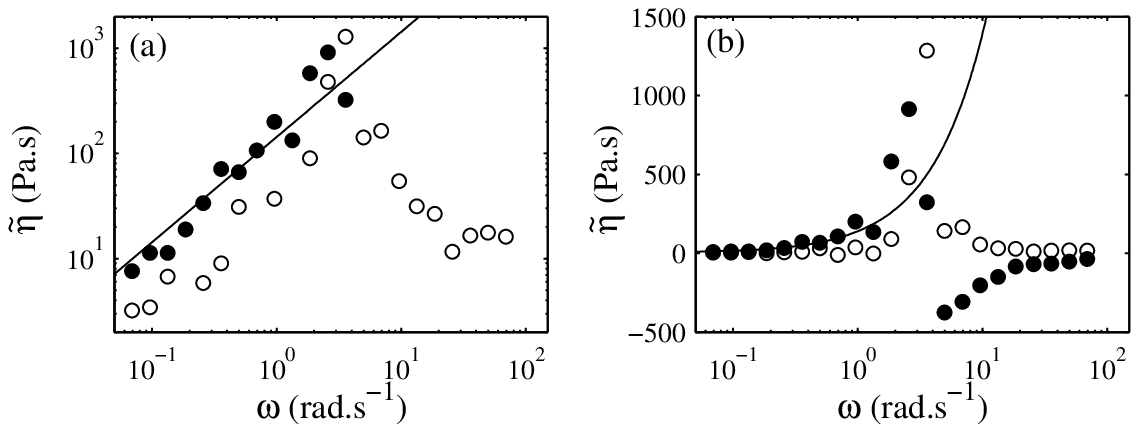}
 \caption{\label{fig5} $\Im(\tilde\eta)$ ($\bullet$) and $\Re(\tilde\eta)$ ($\circ$) versus $\omega$ in (a) logarithmic scales and (b) semi-logarithmic scales. The solid line is the best linear fit of $\Im(\tilde\eta)$ by eq.~(\ref{imag_eta_tilde_approx}) with $\dot\gamma_I=2.2$~s$^{-1}$, $\dot\gamma_N=7.4$~s$^{-1}$, $\sigma_c=100$~Pa, $e=1$~mm, and $c_0=0.1$~mm.s$^{-1}$. The fluid under study is an 8$\%$~wt. CPCl--NaSal solution.}
\end{figure}

In order to use the extrapolation procedure proposed in sect.~\ref{ipp} for eq.~(\ref{eq_detc0}), $\eta_{\| I}^*$ is taken to be the experimental value for the homogeneous fluid obtained closest to the onset of shear banding. The corresponding data are shown in fig.~\ref{fig3c}(a) ($\circ$ symbols, see also the discussion in sect.~\ref{discuss}). We then calculate $1/\tilde\eta=1/\eta_{\|}^*(\alpha_1\rightarrow 0)-1/\eta_{\| I}^*$ for various frequencies $\omega$ ranging from 0.07 to 70~rad.s$^{-1}$. The real and imaginary parts of $\tilde\eta$ are plotted as a function of $\omega$ in fig.~\ref{fig5}.

If eq.~(\ref{eq_annexe_finale2}) holds, one expects
\begin{equation}
\tilde\eta(\omega) = \frac{\sigma_c}{\dot\gamma_N-\dot\gamma_I}\,\frac{i \omega e}{c_0}\left(\frac{\epsilon_{co}}{1-\frac{2ic_0}{\omega R_0}}+2\epsilon_{cp}\right)^{-1}\,.\label{eta_tilde}
\end{equation}
If one further assumes that $2c_0/\omega R_0\ll 1$, then one should find a range of $\omega$ for which $\Im(\tilde\eta)\gg\Re(\tilde\eta)$ and
\begin{equation}
\label{imag_eta_tilde_approx} \Im(\tilde\eta) \simeq \frac{\sigma_c}{\dot\gamma_N-\dot\gamma_I}\,\frac{\omega e}{c_0}\,\frac{1}{1+\epsilon_{cp}}\,.
\end{equation}
Figure~\ref{fig5}(a) shows that $\Im(\tilde\eta)\gg\Re(\tilde\eta)$ for $\omega\simeq 0.07$--3~rad.s$^{-1}$ in the experiment. The best linear fit of $\Im(\tilde\eta)$ vs $\omega$ over this range of frequencies yields $c_0(1+\epsilon_{cp})=0.13\pm 0.05$~mm.s$^{-1}$ so that $c_0=0.1\pm 0.04$~mm.s$^{-1}$. The large uncertainty ($\simeq 40\%$) on the determination of $c_0$ is mainly due to the uncertainty on $\dot\gamma_I$ and $\dot\gamma_N$ and therefore on the calculation of $\alpha_1$. Since $2c_0/\omega R_0\simeq 0.003$--0.1 for $\omega\simeq 0.07$--3~rad.s$^{-1}$, the approximation leading to eq.~(\ref{imag_eta_tilde_approx}) is justified {\it a posteriori}. These results were obtained with $\dot\gamma_I=2.2$~s$^{-1}$, $\dot\gamma_N=7.4$~s$^{-1}$, and $\sigma_c=100$~Pa, which were estimated independently from nonlinear rheology as explained in sect.~\ref{subsec_co}. However, at ``high'' frequencies ($\omega\gtrsim 1$~rad.s$^{-1}$), the terms induced by the curvature of the Mooney-Couette geometry are no longer negligible, so that first-order terms in $e/R_0$ should be taken into account in eq.~(\ref{eq_E_co}). This most probably explains the observation of negative data for $\Im(\tilde\eta)$ in fig.~\ref{fig5}(b).

\begin{figure}
 \includegraphics{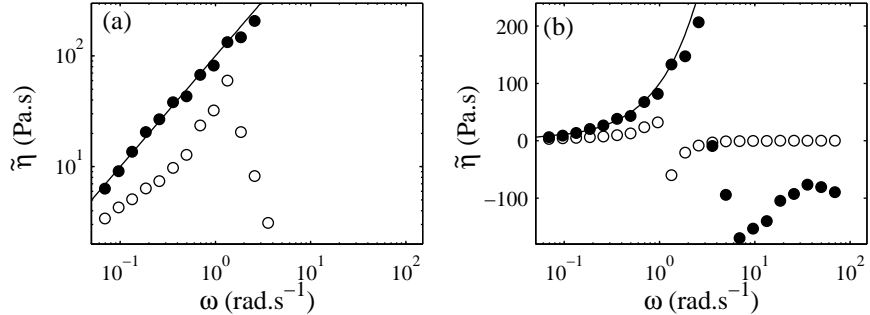}
 \caption{\label{fig5b} $\Im(\tilde\eta)$ ($\bullet$) and $\Re(\tilde\eta)$ ($\circ$) versus $\omega$ in (a) logarithmic scales and (b) semi-logarithmic scales. The solid line is the best linear fit of $\Im(\tilde\eta)$ by eq.~(\ref{imag_eta_tilde_approx}) with $\dot\gamma_I=4.0$~s$^{-1}$, $\dot\gamma_N=6.3$~s$^{-1}$, $\sigma_c=68$~Pa, $e=1$~mm, and $c_0=0.31$~mm.s$^{-1}$. The fluid under study is a 6$\%$~wt. CPCl--NaSal solution.}
\end{figure}

In the last section of this paper we use ultrasonic velocimetry to directly access the dynamics of the interface during superposition experiments and check the validity of the above findings. These experiments were performed on a 6\% wt. CPCl--NaSal solution (due to technical limitations involving the velocimetry setup and the 8\% wt. sample). To allow for a direct comparison with velocimetry experiments, fig.~\ref{fig5b} presents the analysis of superposition rheology measurements performed on the 6\% wt. solution. The results are qualitatively the same as those for the 8\% wt. sample shown in fig.~\ref{fig5}. The estimate for $c_0$ in the 6\% wt. sample is $c_0=0.31\pm 0.15$~mm.s$^{-1}$.

\subsection{Ultrasonic velocimetry during superposition experiments}
\label{usv}
\subsubsection{Velocity profile measurements}

Superposition experiments in the shear banding regime have shown the possibility of characterizing the dynamics of the interface between shear bands using only a standard rheometer. In this section, the above results and model are confirmed using time-resolved local velocity measurements. To access the velocity field we used the ultrasonic velocimetry technique described in \cite{Manneville:2004}. As shown by \cite{Becu:2004} this technique allows one to measure the velocity profile of shear-banding wormlike micelles in the gap of a Couette cell with a temporal resolution of about 1~s and a spatial resolution of about 40~$\mu$m.

\begin{figure}
 \includegraphics{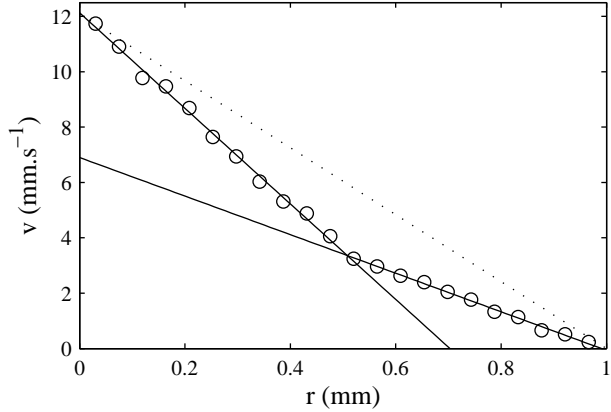}
 \caption{\label{fig6} Velocity profile $v(r)$ ($\circ$) recorded in a $6\%$~wt. CPCl--NaSal solution at steady-state for $\sigma_1=70.5$~Pa (which corresponds to $\alpha_1\simeq 0.5$). The solid lines represent linear fits of the velocity profile in the high- and low-shear bands. Their intersection yields the position $r_c$ of the interface. The dotted line shows the velocity profile for a Newtonian fluid.}
\end{figure}

Figure~\ref{fig6} shows a typical velocity profile $v(r)$ measured in a $6\%$~wt. CPCl--NaSal solution, where $r$ is the distance from the inner rotating cylinder. As explained in \cite{Manneville:2004}, the fluid was seeded with $1\%$~wt. hollow glass spheres (Sphericel, Potters Industries) of mean radius $11.7$~$\mu$m and density $1.1$ in order to provide acoustical scattering. We checked that both linear and nonlinear rheological properties were not significantly affected by the addition of such acoustic contrast agents. Since the velocity profiles are recorded in the Couette part of the Mooney-Couette cell, we shall focus on the model developed in sect.~\ref{subsec_co}. Let us only recall here eq.~(\ref{eq_a2_co}) which gives the complex amplitude $r_2=\alpha_2 e$ of the interface displacement:
\begin{equation}
\label{eq_r2_co}r_2 = \frac{\sigma_2}{\sigma_c}\,\frac{R_0^2}{(R_0+r_1)^2}\,\frac{c_0}{i\omega +\frac{2c_0}{R_0+r_1}}\,.
\end{equation}

In the following, the steady shear stress is fixed to $\sigma_1$ such that $\alpha_1\simeq 0.5$. The steady-state velocity profile of fig.~\ref{fig6} clearly shows two linear parts that separate the gap into two shear bands of equal width where the apparent viscosities differ by a factor of about two. Linear fits in the two shear bands yield the interface position $r_1=\alpha_1 e\simeq 0.5$~mm.

\subsubsection{Measurement of $c_0$ in a transient experiment}

\begin{figure}
\includegraphics{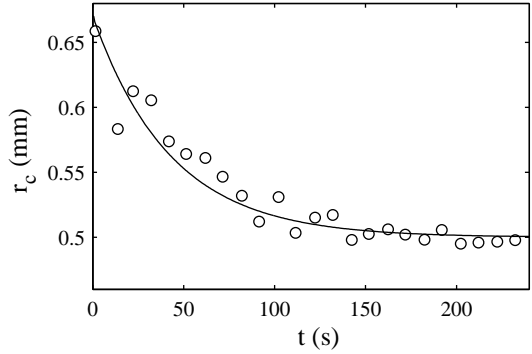}
\caption{\label{fig7} Position of the interface $r_c(t)$ versus time as the external shear stress is reduced from $\sigma_1+\sigma_2=71.5$~Pa to $\sigma_1=70.5$~Pa at $t=0$. The solid line is the best fit by eq.~(\ref{eq_veloc_relax}) with $r_2=0.17$~mm and $c_0=0.28$~mm.s$^{-1}$. The fluid under study is a 6$\%$~wt. CPCl--NaSal solution. }
\end{figure}

In the framework of the model proposed by \cite{Radulescu:1999} the characteristic velocity $c_0$ can be deduced from transient velocity profile measurements. Indeed by suddenly decreasing the shear stress from $\sigma=\sigma_1+\sigma_2$ to $\sigma=\sigma_1$ at time $t=0$ and by measuring the evolution of the velocity profiles $v(r,t)$ in time, we can easily track the interface position $r_c(t)$. Experimentally $\sigma_2$ is fixed such that, by using eq.~(\ref{r2_co}), $r_2(\omega=0)=\alpha_2(\omega=0) e\simeq 0.2$~mm. In the small-gap approximation, the equation for the interface position reads:
\begin{equation}
\frac{1}{c_0}\,\frac{\hbox{\rm d}r_c}{\hbox{\rm d}t}=\frac{\sigma (r_c) -\sigma_c}{\sigma_c}=
\frac{2}{R_0}\,\left(r_1-r_c(t)\right)\,,
\end{equation}
which leads to
\begin{equation}
r_c(t)=r_1+r_2\,e^{-2c_0t/R_0}\,,
\label{eq_veloc_relax}
\end{equation}
where $r_2(\omega=0)$ was simply noted $r_2$. As seen in fig.~\ref{fig7}, the position of the interface $r_c(t)$ is well fitted by eq.~(\ref{eq_veloc_relax}) which yields $c_0=0.28\pm0.03$~mm.s$^{-1}$. Comparing with the results of the superposition measurements shown in fig.~\ref{fig5b}, one finds that both values are in quantitative agreement, which confirms the relevance and the ability of superposition rheology to extract dynamical information in the shear banding regime. Of course the uncertainty on $c_0$ given by time-resolved velocimetry is much less than that of the superposition method (but at the cost of using a more involved technique and processing a large amount of ultrasonic data).

The present analysis of the velocity measurements also neglects the first two stages of the band dynamics during the transient, namely low-shear band destabilization and interface reconstruction, as evidenced by \cite{Radulescu:2003}. These initial stages were shown to occur in typically 2~s which is of the order of the temporal resolution of our velocimetry experiments. Thus we only focus on the last dynamical step called ``interface travel'' in \cite{Radulescu:2003}. In particular, due to the existence of two early relaxation stages, one may argue that the initial position is ill-defined and that the interface position after reconstruction may significantly differ from $r_1+r_2$. This is the reason why $r_2$ was actually left as a free parameter in eq.~(\ref{eq_veloc_relax}).

\subsubsection{Validation of the model for superposition experiments}

\begin{figure}
 \includegraphics{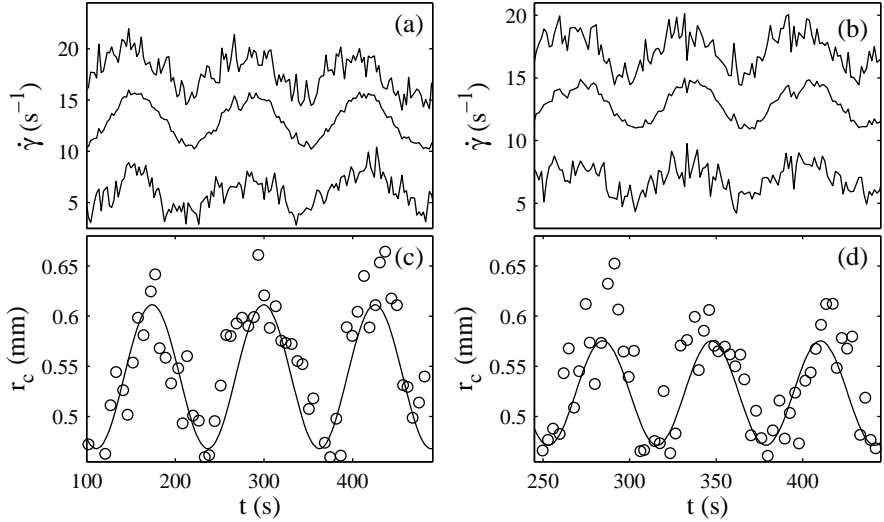}
 \caption{\label{fig8} (a) Local shear rates versus time in the nematic band (top) and isotropic band (bottom) along with the global shear rate recorded by the rheometer (middle) for $\omega=0.05$~rad.s$^{-1}$ and (b) $\omega=0.1$~rad.s$^{-1}$. Position of the interface $r_c(t)$ versus time for (c) $\omega=0.05$~rad.s$^{-1}$ and (d) $\omega=0.1$~rad.s$^{-1}$. The solid lines are the best fits by sine functions. The fluid under study is a 6$\%$~wt. CPCl--NaSal solution submitted to stress oscillations of amplitude $\sigma_2=1$~Pa around the mean value $\sigma_1=70.5$~Pa.}
\end{figure}

Now turning to the superposition experiment, we add an oscillatory shear stress of amplitude $\sigma_2$ to a steady shear stress $\sigma_1$, and follow the position of the interface in time for various frequencies. Figure~\ref{fig8} presents the measurements of the local shear rates in the two bands and of the interface position $r_c(t)$ versus time for two different frequencies. As expected these various quantities oscillate in time and, in spite of some experimental scatter, fitting $r_c(t)$ by sine functions for various frequencies yields a good estimate for the amplitude $r_2(\omega)=\alpha_2(\omega) e$.

\begin{figure}
 \includegraphics{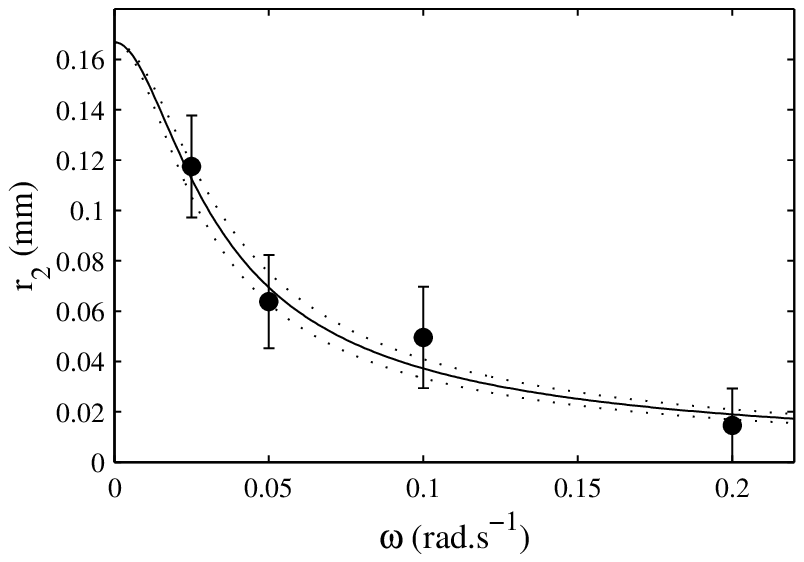}
 \caption{\label{fig9} Amplitude of the interface displacement $\mid r_2\mid$ versus $\omega$. The solid line represents the prediction of eq.~(\ref{eq_r2_co}) where all the parameters $\sigma_2=1$~Pa, $\sigma_c=68$~Pa, $r_1=0.5$~mm, and $c_0=0.28$~mm.s$^{-1}$ are known independently. The dotted lines were computed using $c_0=0.25$ (lower curve) and $c_0=0.31$~mm.s$^{-1}$ (upper curve) in eq.~(\ref{eq_r2_co}). They illustrate the sensitivity of the prediction to a 10\% variation in $c_0$, which corresponds to the experimental uncertainty on the fit of fig.~\ref{fig7}. The fluid under study is a 6$\%$~wt. CPCl--NaSal solution.}
\end{figure}

The dots ($\bullet$) in Fig.~\ref{fig9} show the amplitude $\mid r_2\mid$ of the interface oscillations inferred from ultrasonic velocimetry for four different frequencies, while the solid line is calculated using eq.~(\ref{eq_r2_co}) with the value $c_0=0.28$~mm.s$^{-1}$ obtained from the transient experiment. The quantitative agreement between the experimental data and the calculated prediction confirms the generality of eq.~(\ref{eq_r2_co}) and provides strong support for the model developed in sect.~\ref{model}. Let us emphasize that in the present case the prediction for $r_2(\omega)$ is obtained without any free parameter since $\sigma_2$, $r_1$, $R_0$, and $e$ are known experimentally and $\sigma_c$ is found by nonlinear rheology.

\subsection{Discussion and perspectives}
\label{discuss}

Our main result is that superposition rheology can be used to infer conclusive information on the dynamics of wormlike micelles in the shear banding regime. In particular, superposition measurements lead to an estimate of the velocity $c_0$ which characterizes the dynamics of the interface between shear bands. The present uncertainty on the estimation of $c_0$ through superposition rheology alone is of the order of $\pm40\%$. In our opinion this relatively large uncertainty is due to the use of a Mooney-Couette cell and to the subsequent approximations needed to process the superposition data in order to recover $c_0$. Experiments in the cone-and-plate geometry under controlled shear rate should be simpler to process and should provide a better accuracy on $c_0$. To minimize boundary effects in the concentric cylinder geometry, one could also avoid the use of a Mooney-Couette cell by trapping an air bubble below the inner cylinder.

Let us now discuss the value of $c_0$ found from the superposition experiments reported above. According to \cite{Radulescu:2003}, $c_0$ is linked to the diffusion coefficient ${\cal D}$ of the stress across the streamlines, a central parameter in recent theoretical approaches of shear banding \cite{Olmsted:2000}. More precisely, one has
\begin{equation}
\label{eq_radul_1} \left.\frac{d\sigma}{dc}\right|_{\sigma=\sigma_c}=\frac{\sigma_c}{c_0}=K G_0\sqrt{\frac{\tau_1}{{\cal D}}}\,,
\end{equation}
where $c=e{\rm d}\alpha/\hbox{\rm d}t$ is the velocity of the interface, $K$ is a dimensionless parameter that depends on the constitutive model, $G_0$ is the plateau modulus, and $\tau_1$ the main relaxation time already introduced in sect.~\ref{convrheol}. Following \cite{Radulescu:2003}, we take $K G_0/\eta^*_{\|}(0,\dot\gamma_I)\dot\gamma_I=0.3$. From the nonlinear rheological measurements of sect.~\ref{convrheol} and using eq.~(\ref{eta0}), we find $\eta^*_{\|}(0,\dot\gamma_I)\dot\gamma_I\simeq 0.09 \sigma_c$. With $c_0=0.1$~mm.s$^{-1}$ and $\tau_1=0.87$~s,  eq.~(\ref{eq_radul_1}) yields ${\cal D}\simeq 6.3~10^{-12}$~m$^{2}$.s$^{-1}$ for the 8\% wt. CPCl--NaSal solution. This corresponds to a stress correlation length $\zeta=\sqrt{{\cal D}\tau_1}\simeq 2.3~\mu$m, which is much larger than the mesh size $\xi\sim (kT/G_0)^{1/3}\simeq 30$~nm of our system. Let us emphasize the fact that ultrasonic velocimetry has provided the same order of magnitude for $c_0$ at a slightly lower surfactant concentration but with a much better accuracy ($c_0=0.28\pm0.03$~mm.s$^{-1}$ for a 6\% wt. CPCl--NaSal solution), which confirms that the stress diffusion coefficient estimated from eq.~(\ref{eq_radul_1}) should be in the range $10^{-12}$--10$^{-11}$~m$^{2}$.s$^{-1}$ in the system under study.

Such a value of ${\cal D}$ differs by two orders of magnitude from the stress diffusion coefficient inferred from transient rheo-optical measurements by \cite{Radulescu:2003} in various wormlike micelle solutions with 0.3~M CTAB (${\cal D}\simeq 1.2-7.2~10^{-14}$~m$^{2}$.s$^{-1}$). Consequently our stress correlation length is about twenty times larger than the estimate found by \cite{Radulescu:2003}, $\zeta\simeq 100$~nm, which was comparable to the mesh size of their micellar network ($\xi\simeq 26$~nm). Since we used the CPCl--NaSal system rather than CTAB solutions such a difference may not be too unexpected. For instance the viscosity difference between the two coexisting phases, and hence the width of the stress plateau, is much smaller in our case ($\tau(\dot\gamma_N-\dot\gamma_I)\simeq 4.5$) than in the experiments of \cite{Radulescu:2003} ($\tau(\dot\gamma_N-\dot\gamma_I)\simeq 19$--80). Moreover the value of $K$ is not only model-dependent but could also vary with the average shear rate \cite{Dhont:1999}. Estimates of ${\cal D}$ inferred from eq.~(\ref{eq_radul_1}) should thus probably be taken with care. In any case, superposition experiments in CTAB solutions, where wider stress plateaus and a better precision on their limits should make the determination of $c_0$ more accurate, would be very useful in order to confirm the values of ${\cal D}$ found by \cite{Radulescu:2003} in this system.

Besides extracting a value for ${\cal D}$, superposition rheology in the shear banding regime would be even more interesting if it could provide some information on the rheological behaviour of the nematic phase as suggested in sect.~\ref{ipp}. Indeed information about the structure and dynamics of the shear-induced, oriented phase is often tricky to derive from conventional measurements due to the slope in the flow curve that results from curvature and due to instabilities that occur on the high-shear branch. Here the dynamical behavior $\eta_{\| N}^*(\omega)$ of the shear-induced phase may be recovered by considering the slopes of the linear fits of $1/\eta_{\|}^*$ in fig.~\ref{fig4} which are equal to $1/\eta_{\| I}^*-1/\eta_{\| N}^*$ according to eq.~(\ref{eq_annexe_finale2}). The reconstructed $\eta_{\|N}^*(\omega)$ data are presented in fig.~\ref{fig3c}(b), where they are compared to superposition data measured at the beginning of the high-shear branch of the flow curve. The fact that the experimental data are systematically lower than the reconstructed data can be easily explained by the distance from the experimental shear rate ($\dot\gamma=11.1$~s$^{-1}$) to the upper limit of the stress plateau ($\dot\gamma_N\simeq 7.4$~s$^{-1}$). Moreover, as shown in fig.~\ref{fig3c}(a), which compares experimental data recorded just below $\dot\gamma_I$ and the $\eta_{\|I}^*(\omega)$ data reconstructed using the $\alpha_1=0$ limit and $c_0=0.1$~mm.s$^{-1}$ in eq.~(\ref{eq_annexe_finale2}), superposition measurements at the onset of shear banding also yield a very good approximation of the complex viscosity $\eta_{\| I}^*(\omega)$ close to the beginning of the plateau. This allowed us to check the consistency of our fitting procedure and to confirm that superposition rheology provides useful quantitative information on the dynamical behaviours of both the entangled and the oriented states. A deeper analysis and modelling of such behaviours are left for future work.

\begin{figure}
 \includegraphics{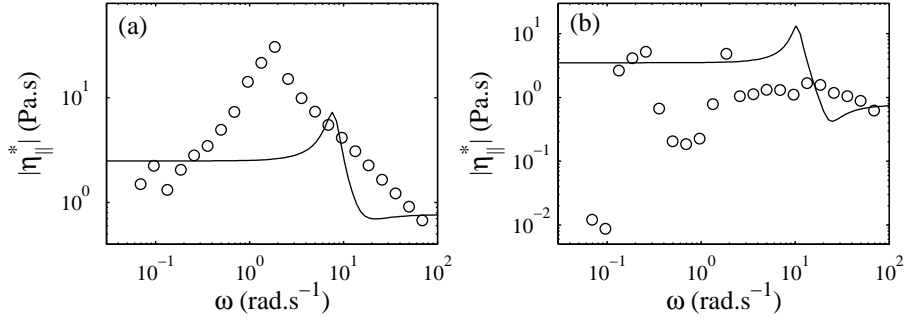}
 \caption{\label{fig3d} Superposition rheology of an 8$\%$~wt. CPCl--NaSal solution in the high-shear regime: $\mid\eta_{\| }^*(\omega,\dot\gamma_1)\mid$ versus $\omega$ for (a) $\dot\gamma_1=11.1$ and (b) 15.0~s$^{-1}$. The solid lines correspond to an Oldroyd-B fluid (eq.~(\ref{eq_annexe_oldroy_eta2})) with $\eta_0=122$~Pa, $s_1=0.59$~s, $s_2=0.13$~s, $\tau_1=0.87$~s, and $\tau_2=0.60$~ms.}
\end{figure}

For the sake of completeness, fig.~\ref{fig3d} shows superposition data obtained on the high-shear branch of the flow curve. Although a simple interpretation of fig.~\ref{fig3d}(b) may not be possible due to the occurrence of flow instabilities for $\dot\gamma\gtrsim 15$~s$^{-1}$, these data clearly show that the dynamical behaviour of the shear-induced phase totally differs from the initial Oldroyd-B behaviour of the weakly oriented, entangled phase. Such information may turn out to be crucial for the modelling of shear banding since the exact behaviour of the fluid at the limits of the stress plateau is usually unknown.

Finally the influence of normal stresses or flow-concentration coupling in superposition experiments and the way to include them in a model also constitute directions for further research.

\section*{Conclusion}

In this paper we have shown that superposition rheology constitutes a useful tool to access the dynamics of a shear-banded flow. A two-fluid semi-phenomenological model was proposed based on the simplest shear banding scenario. This model was shown to provide a good description of the oscillations of the interface between shear bands in CPCl--NaSal wormlike micelle solutions sheared in the Mooney-Couette geometry. In particular an estimate for the stress diffusion coefficient ${\cal D}$ was reported for the first time in the CPCl--NaSal micellar system, whose value was shown to be significantly larger than that reported for CTAB systems. Independent measurements of the interface dynamics through local velocimetry experiments nicely corroborated our model without any free parameter. We have shown, however, that a more accurate determination of the characteristic velocity requires a simpler and better controlled geometry. Further experiments, {\it e.g.} under controlled shear rate in the cone-and-plate geometry, should allow one to probe even more precisely the dynamics of the shear bands using only a standard rheometer and to infer important information on the dynamical behaviours of the two coexisting phases. The formalisms to use for these experiments are also supplied in the present work.

\begin{acknowledgments}
The authors wish to thank J.~Teisseire, A.~Colin, F.~Molino, and C.~Gay for fruitful discussions as well as the ``Cellule Instrumentation'' of CRPP for technical advice and design of the experiment. This work was cofunded by CNRS, R{\'e}gion Aquitaine, and the SoftComp Network of Excellence, a project of the European Commission developed under the sixth Framework Programme.

\end{acknowledgments}

\appendix
\section{Two-fluid calculations in experimental geometries}

In this appendix the detailed calculations for two-fluid superposition rheology are presented in the standard geometries used in the experiments namely cone-and-plate, Couette, and Mooney-Couette geometries.

\subsection{Cone-and-plate geometry}
\label{subsec_cp}

Let us first consider a cone-and-plate geometry of angle $\beta\ll 1$ and maximum radius $R_{0}$. In such a geometry and in a homogeneous fluid, the shear rate can be considered as constant throughout the sample. In the shear banding regime, $\dot\gamma_1$ and $\alpha_1$ are still linked by the lever rule (\ref{eq_lr2}) so that eq.~(\ref{eta_ipp}) remains valid. Moreover, at a given distance $r$ from the axis of the cone, the system is equivalent to an infinite parallel plate geometry of gap $e=r\tan\beta$ for which the shear stress is
\begin{equation}
\sigma (r) =\sigma_c + \eta_{\| }^*(r)\dot\gamma_2\,e^{i\omega t}\,,
\label{sigR}
\end{equation}
where $\eta_{\| }^*(r)$ is computed from eq.~(\ref{eq_annexe_finale}) by setting $e=r\tan\beta$. One can then calculate the total stress exerted on the cone from
\begin{equation}
\sigma = \frac{2}{ R_{0}^2} \int _0^{R_{0}} \sigma (r)\, r\,\hbox{\rm d}r \,.
\label{sigtot}
\end{equation}
In analogy with eqs.~(\ref{visc_lr}) and (\ref{visc_dyn}) let us define the two characteristic viscosities
\begin{eqnarray}
\frac{1}{\eta_{L}}&=&\frac{1-\alpha_1}{\eta^*_{\| I}}+\frac{\alpha_1}{\eta^*_{\| N}}\,,\label{visc_lr_cp}\\
\frac{1}{\eta_{D}}&=&\frac{ \dot\gamma_N-\dot\gamma_I}{\sigma_c}\, \frac{c_0}{i \omega R_{0} \tan\beta}\,,\label{visc_dyn_cp}
\end{eqnarray}
so that $\eta_{D}$ corresponds to the dynamical term $\eta_{D_\infty}$ of the complex viscosity (eq.~(\ref{visc_dyn})) with $e=R_{0}\tan\beta$. With these notations, inserting eq.~(\ref{sigR}) into eq.~(\ref{sigtot}) leads to
$\sigma=\sigma_c+\eta_{\| }^*\dot\gamma_2\,e^{i\omega t}$ where
\begin{equation}
\label{eq_E_cp}\eta_{\| }^*  = \eta_{L} \left[ 1 -2\frac{\eta_{L}}{\eta_{D}}+2\frac{\eta_{L}^2}{\eta_{D}^2} \ln \left( 1+\frac{\eta_{D}}{\eta_{L}} \right) \right]\,.
\end{equation}

The above expression for $\eta_{\| }^*$ in the cone-and-plate geometry clearly differs from eq.~(\ref{eq_annexe_finale}) obtained for infinite parallel plates. In particular a linear fit of $1/\eta_{\| }^*$ vs $\alpha_1$ does not seem relevant. As discussed in sect.~\ref{ipp}, one could still use eq.~(\ref{eq_E_cp}) to fit $\eta_{\| }^*$ with two free parameters $\eta^*_{\| N}(\omega)$ and $\eta_{D}(\omega)$. Such a procedure would provide an estimate for $\eta_{D}(\omega)$ and therefore $c_0$. Another way to proceed is to notice that in our experiments $\eta_{D} > \eta_{L}$, so that
\begin{equation}
\frac{1}{\eta_{\| }^*} \simeq \frac{1}{\eta_{L}}+\frac{2}{\eta_{D}}\,,
\label{eq_E_cp2}
\end{equation}
which is equivalent to eq.~(\ref{eq_annexe_finale}) with $2e=R_{0}\tan\beta$. In this case the linear regression of $1/\eta_{\| }^*$ may also lead to a good approximation of $c_0$.

\subsection{Couette geometry}
\label{subsec_co}

Let us now consider a concentric cylinder geometry (Couette geometry) where the inner cylinder of radius $R_0$ is rotating while the outer cylinder of radius $R_1$ remains fixed. This choice is made to be consistent with the experimental section but our model can easily be adapted to any rotational configuration of the two cylinders. The gap between the rotor and the stator is $e=R_1-R_0$. In the Couette geometry the shear stress is {\it not} homogeneous throughout the whole cell. Under the steady-state approximation already discussed in sect.~\ref{ipp}, the shear stress depends on the distance $r$ from the inner cylinder as
\begin{equation}
\label{eqco_sig} \sigma (r)=\frac{\sigma_1 +\sigma_2 e^{i\omega t}}{\left(1+\frac{r}{R_0}\right)^2}\,,
\end{equation}
so that $\sigma_1 +\sigma_2 e^{i\omega t}=\sigma(0)$ corresponds to the shear stress at the inner cylinder. Note that the rheometer may rather indicate ``average'' shear stresses for $\sigma_1$ and $\sigma_2$ measured in superposition experiments. Since these stresses only differ from the values at the inner cylinder by a geometrical factor of order 1 and since this factor also depends on the way the average is defined, we shall leave out this complication and stick with $\sigma_1$ and $\sigma_2$ as the values at the inner wall.

In the simple shear banding scenario described in the introduction, the inhomogeneity of $\sigma$ ensures that only two shear bands separated by a single interface coexist in the gap. More precisely the shear-induced transition occurs when there exists $0\leq r_c\leq e$ such that $\sigma(r_c)=\sigma_c$. For $r>r_c$, $\sigma(r)<\sigma_c$ so the fluid remains entangled and in the high-viscosity state, while for $r<r_c$, $\sigma(r)>\sigma_c$ and the fluid is in the shear-induced low-viscosity state. Another consequence of the stress inhomogeneity is that the stress plateau is {\it not} flat \cite{Radulescu:2000,Salmon:2003}. Indeed the shear-induced state first appears when $\sigma(0)=\sigma_c$ and fills the whole cell when $\sigma(0)=\sigma_c(1+e/R_0)^2$. When $e/R_0\ll 1$, this leads to a linear $\sigma$ vs $\dot\gamma$ curve with slope $\hbox{\rm d}\sigma/\hbox{\rm d}\dot\gamma=2e\sigma_c/R_0(\dot\gamma_I-\dot\gamma_N)$. In the case of the experimental data shown in fig.~\ref{fig2}, $e/R_0\simeq 0.04$ and the shear stress is indeed seen to increase linearly in the shear banding regime. However, as already noted, the high-shear branch of the flow curve is hardly distinguishable from the ``stress plateau.'' Still we can take advantage of the existence of a tilted plateau to estimate $\dot\gamma_N$. Fitting the flow curve at high shear rates by a Bingham fluid $\sigma=\sigma_B+\eta_B\dot\gamma$ (as suggested by \cite{Salmon:2003}) and looking for the shear rate corresponding to $\sigma=\sigma_c=100$~Pa yields $\dot\gamma_N=7.4\pm 0.4$~s$^{-1}$ (see dashed line in fig.~\ref{fig2}).

Thus, from eq.~(\ref{eqco_sig}), it is required that $\sigma_c <\sigma_1 \pm \sigma_2 <\sigma_c(1+e/R_0)^2$ for a superposition experiment to be performed in the shear banding regime at all times. Let us define $r_1$ such that $\sigma_c=\sigma_1/(1+r_1/R_0)^{2}$ and $r_c(t)$ the position of the interface at time $t$. The model proposed by \cite{Radulescu:1999} implies that
\begin{equation}
\frac{1}{c_0}\,\frac{\hbox{\rm d}r_c}{\hbox{\rm d}t}=\frac{\sigma (r_c) -\sigma_c}{\sigma_c}=
\,\left( 1+\frac{\sigma_2}{\sigma_1}\,e^{i\omega t } \right) \left( \frac{R_0+r_1}{R_0+r_c} \right)^2-1\,.
\label{posint}
\end{equation}
In the linear response, eq.~(\ref{posint}) leads to $r_c(t)=r_1+r_2\exp(i\omega t)$, where
\begin{equation}
r_2=\frac{\sigma_2}{\sigma_1}\,\frac{c_0}{i\omega +\frac{2c_0}{R_0+r_1}}\,.
\label{r2_co}
\end{equation}
Since $r_c(t)=\alpha(t) e$, one gets $\alpha(t)=\alpha_1+\alpha_2\exp(i\omega t)$ with
\begin{eqnarray}
\label{eq_alp1_co} \alpha_1 &=&\frac{r_1}{e}=\frac{R_0}{e} \left( \sqrt{\frac{\sigma_1}{\sigma_c}}-1 \right)\,,\\
\label{eq_a2_co} \alpha_2 &=&  \frac{r_2}{e}=\frac{\sigma_2}{\sigma_c}\,\frac{R_0^2}{(R_0+r_1)^2}\,\frac{c_0}{i\omega e+\frac{2c_0e}{R_0+r_1}}\,.
\end{eqnarray}
This last equation is tested experimentally through velocity profile measurements in sect.~\ref{usv}.

Once the interface motion is known from eqs.~(\ref{eq_alp1_co}) and (\ref{eq_a2_co}), one can go back to the apparent shear rate, {\it i.e.} the shear rate averaged over the whole sample
\begin{eqnarray}
\dot\gamma(t) = \dot\gamma_1+ \dot\gamma_2\,e^{i\omega t } &=& \int_0 ^{r_c(t)} \left( \frac{\sigma_1 (r)}{\eta_{N}(\sigma_1 (r))}+ \frac{\sigma_2 (r)\,e^{i\omega t }}{\eta^*_{\| N}(\omega,\sigma_1 (r))} \right) \frac{dr}{e}\nonumber\\
&+& \int_{r_c(t)} ^{e} \left( \frac{\sigma_1 (r)}{\eta_{I}(\sigma_1 (r))}+ \frac{\sigma_2 (r)\,e^{i\omega t }}{\eta^*_{\| I}(\omega,\sigma_1 (r))} \right) \frac{dr}{e}\,,
\label{gaminteg}
\end{eqnarray}
where $\sigma_1 (r)=\sigma_1/(1+r/R_0)^2$ and $\sigma_2 (r)=\sigma_2/(1+r/R_0)^2$. Since our superposition experiments are performed under controlled stress, we have noted the viscosities $\eta_I$, $\eta_N$, $\eta^*_{\| I}$, and $\eta^*_{\| N}$ as functions of the local steady shear stress $\sigma_1(r)$. In principle, knowing the different viscosities (from experimental measurements or extrapolated data as mentioned in sect.~\ref{ipp}), eqs.~(\ref{r2_co}) and (\ref{gaminteg}) allow one to solve for $\dot\gamma_1$ and $\dot\gamma_2$ and thus to find $\eta^*_{\|}=\sigma_2/\dot\gamma_2$.

In order to get an explicit form for $\eta^*_{\|}$ that we may compare to eq.~(\ref{eq_annexe_finale}), we shall assume that the small-gap approximation $e \ll R_0$ holds, which is almost always the case in standard experiments in the Couette geometry. In that case, expanding eq.~(\ref{gaminteg}) to first-order in $e/R_0$ and looking for the constant terms leads to
\begin{eqnarray}
\label{eq_e_co}
& &\frac{1}{\eta} = \frac{\dot\gamma_1}{\sigma_1} = \frac{1-\alpha_1}{\eta_{I}(\sigma_c )}+\frac{\alpha_1}{\eta_{N} (\sigma_c )}\\ \nonumber
&+& \frac{e}{R_0}\left[
 \frac{1}{\eta_{I}(\sigma_c )} \left( (1-\alpha_1 )^2\frac{ \sigma_c}{\eta_{I}(\sigma_c )} \frac{\partial \eta_{I} }{\partial \sigma} |_{\sigma_c} - 1+\alpha_1^2 \right)
 - \frac{\alpha_1 ^2}{\eta_{N}(\sigma_c )} \left( \frac{\sigma_c}{\eta_{N}(\sigma_c )} \frac{\partial \eta_{N} }{\partial \sigma} |_{\sigma_c} + 1 \right)
\right]\,.
\end{eqnarray}
This yields the apparent viscosity $\eta$ indicated by the rheometer in the shear banding regime (up to some multiplicative factor of order 1 that depends on whether the rheometer actually indicates the shear stress at the inner wall or some average shear stress, as already mentioned above). Note the first order correction in $e/R_0$ to the case of simple shear given by eq.~(\ref{eta_ipp}). By looking for the terms proportional to $\exp(i\omega t)$ in the first-order expansion of eq.~(\ref{gaminteg}), one finds
\begin{eqnarray}
\label{eq_E_co}
\frac{1}{\eta^*_{\|}} = \frac{\dot\gamma_2}{\sigma_2} = \frac{1}{\eta_{L}} + \frac{1}{\eta_{D}} + \frac{1}{\eta_{\partial}} \,,
\end{eqnarray}
with
\begin{eqnarray}
\frac{1}{\eta_{L}} &=&\frac{1-\alpha_1}{\eta^*_{\| I}}+ \frac{\alpha_1 }{\eta^*_{\| N}}\,, \\
\frac{1}{\eta_{D}} &=&  \frac{ \dot\gamma_N-\dot\gamma_I}{\sigma_c}\, \frac{c_0}{i\omega e +\frac{2c_0 e}{R_0}}
\left[ 1+\alpha_1\frac{e}{R_0} \left(  \frac{1}{i\omega e +\frac{2 c_0 e}{R_0}}-2 \right) \right]
\simeq \frac{1}{\eta_{D_\infty}}\,\frac{1}{1-\frac{2ic_0}{\omega R_0}}\, , \label{etaD_co}\\
\frac{1}{\eta_{\partial}}&=&\frac{e}{R_0} \left[
\frac{1}{\eta^*_{\| I}} \left( (1-\alpha_1 )^2 \frac{\sigma_c}{\eta^*_{\| I}} \frac{\partial \eta^*_{\| I}}{\partial \sigma}|_{\sigma_c} -1+\alpha_1^2 \right)
-\frac{\alpha_1 ^2}{\eta^*_{\| N}} \left( \frac{\sigma_c}{\eta^*_{\| N}} \frac{\partial \eta^*_{\| N}}{\partial \sigma}|_{\sigma_c} +1 \right) \right]\,,\label{etad_co}
\end{eqnarray}
where we have dropped the dependence on $\sigma_c$ of the various viscosities for the sake of clarity. Equation (\ref{eq_E_co}) generalizes eq.~(\ref{eq_annexe_finale}) to the case of a small-gap Couette geometry and shows that $\eta_{\| }^*$ now involves three terms: the lever rule $\eta_{L}$, the dynamical component $\eta_{D}$ that arises from the motion of the interface, and $\eta_{\partial}$ a first-order correction to $\eta_{L}$ similar to that found in eq.~(\ref{eq_e_co}) and linked to the stress inhomogeneity. Keeping in mind that the various viscosities in eq.~(\ref{eq_annexe_finale}) are taken at $\sigma_1=\sigma_c$, the case of two infinite parallel plates is easily recovered from eqs.~(\ref{eq_e_co})--(\ref{etad_co}) when $R_0\rightarrow \infty$. 

Therefore, in a small-gap Couette geometry, $1/\eta_{\| }^*$ is a second-order polynomial in $\alpha_1$ whose value for $\alpha_1\rightarrow 0$ is
\begin{equation}
\lim_{\alpha_1\rightarrow 0}\,\frac{1}{\eta_{\| }^*} =\frac{1}{\eta^*_{\| I}}+ \frac{ \dot\gamma_N-\dot\gamma_I}{\sigma_c}\,\frac{c_0}{i\omega e +\frac{2c_0 e}{R_0}}\,.\label{eq_detc0_co}
\end{equation}
This is very similar to eq.~(\ref{eq_detc0}) so that the same data analysis should lead to the measurement of $\eta_D(\omega)$ and to an experimental determination of $c_0$.

\subsection{Mooney-Couette geometry}
\label{subsec_mc}

Experimentally, in order to minimize boundary effects due to the finite height of the cylinders, one often uses a composite geometry, called the Mooney-Couette geometry, made of a Couette cell of gap $e$ with a cone-shaped bottom such that $e=R_0\tan\beta\simeq R_0\beta$. In a Newtonian fluid and in the small-gap approximation, this geometry ensures that the shear rate remains constant over the whole sample.

Using the results obtained in sect.~\ref{subsec_cp} and \ref{subsec_co}, one can easily construct a model for superposition experiments in the Mooney-Couette geometry of height $h$ by considering the proportions $\epsilon_{co}=(1+R_0/2h)^{-1}$ and $\epsilon_{cp}=1-\epsilon_{co}$ of the surface respectively covered by the Couette ($co$) and by the cone-and-plate ($cp$) geometries relative to the total surface. The total shear stress is then simply given by $\sigma =\epsilon_{co} \sigma_{co}+\epsilon_{cp}\, \sigma_{cp}$, which yields
\begin{eqnarray}
\label{eq_e_mc2} \eta &=&\epsilon_{co} \eta_{co}+\epsilon_{cp}\,\eta_{cp}\,,\\
\label{eq_e_mc} \eta_{\| }^* &=&\epsilon_{ co} \eta_{\| co}^*+\epsilon_{cp}\,\eta_{\| cp }^*\,,
\end{eqnarray}
where $\eta_{cp}$ and $\eta_{\| cp}^*$ are given by eqs.~(\ref{eta_ipp}) and (\ref{eq_E_cp2}), and $\eta_{co}$ and $\eta_{\| co}^*$ by eqs.~(\ref{eq_e_co}) and (\ref{eq_E_co}). To close this set of equations, one has to specify the values of $\alpha_1$ in the two parts of the geometry. Since the shear rate is perfectly homogeneous in the cone-and-plate, the steady component of the shear stress acting on the cone is $\sigma_{1cp}=\sigma_c$ so that the steady component of the shear stress acting on the inner cylinder is $\sigma_{1co} =(\sigma_1 -\epsilon_{cp}\,\sigma_c)/\epsilon_{co}$.
Thus the local proportions of shear-induced structure $\alpha_{1cp}$ and $\alpha_{1co}$ are given by
\begin{eqnarray}
\label{eq_a1_co} \alpha_{1co}&=& \frac{R_0}{e} \left( \sqrt{\frac{\sigma_1 -\epsilon_{cp}\,\sigma_c}{\epsilon_{co} \sigma_c}}-1 \right)\,,\\
\label{eq_a1_cp} \alpha_{1cp}&=&\frac{\dot\gamma_1 -\dot\gamma_I}{\dot\gamma_N -\dot\gamma_I}\,.
\end{eqnarray}
In the limit $e/R_0 \ll 1$ one can define an effective $\alpha_1$ for the whole cell:
\begin{equation}
\label{eq_a1_mc} \alpha_1 =\epsilon_{co}\alpha_{1co}+\epsilon_{cp}\,\alpha_{1cp}=\epsilon_{co} \frac{R_0}{e} \left( \sqrt{\frac{\sigma_1 -\epsilon_{cp}\,\sigma_c}{\epsilon_{co} \sigma_c}}-1 \right) +\epsilon_{cp}\, \frac{\dot\gamma_1 -\dot\gamma_I}{\dot\gamma_N -\dot\gamma_I}\,.
\end{equation}

With eqs.~(\ref{eq_e_mc}), (\ref{eq_E_cp2}), (\ref{eq_E_co}), (\ref{eq_a1_co}), and (\ref{eq_a1_cp}), one can in principle determine the characteristic velocity $c_0$ and the dynamical behaviours of the two coexisting phases $\eta_{\| I}^*$ and $\eta_{\| N}^*$ by fitting $\eta_{\|}^*$ using eq.~(\ref{eq_e_mc}) at a fixed $\omega$. However, in practice, such a fit requires to know precisely $\sigma_c$, $\dot\gamma_I$, and $\dot\gamma_N$ together with $\dot\gamma_1$ and $\eta_{\|}^*$ for at least four different values of $\sigma_1$. As already pointed out, $\sigma_c$, $\dot\gamma_I$, and $\dot\gamma_N$ may be difficult to access and, in a curved geometry, are known to within $10\%$ at best. Therefore the complexity of the fitting procedure along with the high number of unknowns prevent us to fit experimental data to the full model described above. Moreover the simple data analysis proposed in sect.~\ref{ipp} and based on an extrapolation to $\alpha_1=0$ (in order to remove the dependence on the unknown viscosity $\eta_{\|N}^*$) is no longer possible in the Mooney-Couette geometry since $\alpha_{1cp}$ and $\alpha_{1co}$ do not go to zero for the same $\dot\gamma_1$ or $\sigma_1$. Nevertheless, in sect.~\ref{subsec_geom}, it is shown that eq.~(\ref{eq_annexe_finale}) along with $\alpha_1$ calculated from eq.~(\ref{eq_a1_mc}) may still allow us to estimate $c_0$.

\end{document}